\documentclass[11pt,a4paper]{article}
\pdfoutput=1

\usepackage{jcappub}
\usepackage{amsmath}

\newcommand{\dd}{\mathrm{d}}
\renewcommand{\vec}[1]{{\bf #1}}

\title{A new analytic solution for 2nd-order Fermi acceleration}

\author{Philipp Mertsch} \affiliation{Rudolf Peierls Centre for
  Theoretical Physics, University of Oxford, \\ 1 Keble Road, Oxford
  OX1 3NP, UK} \emailAdd{p.mertsch1@physics.ox.ac.uk}

\abstract{A new analytic solution for 2nd-order Fermi acceleration is
  presented. In particular, we consider \emph{time-dependent} rates
  for stochastic acceleration, diffusive and convective escape as well
  as adiabatic losses. The power law index $q$ of the turbulence
  spectrum is unconstrained and can therefore account for Kolmogorov
  ($q = 5/3$) and Kraichnan ($q = 3/2$) turbulence, Bohm diffusion ($q
  = 1$) as well as the hard-sphere approximation ($q = 2$). This
  considerably improves beyond solutions known to date and will prove
  a useful tool for more realistic modelling of 2nd-order Fermi
  acceleration in a variety of astrophysical environments.}


\begin{document}
\maketitle
\flushbottom

\section{Introduction}

Stochastic acceleration of relativistic particles by plasma wave
turbulence -- a 2nd-order Fermi process~\cite{Fermi:1949ee} -- is
dominating the production of non-thermal particle distributions in a
variety of astrophysical environments. Radio
galaxies~\cite{Lacombe:1977aa,Eilek:1979aa,Achterberg:1979aa,Hardcastle:2008jw,O'Sullivan:2009sc,Tramacere:2011qw},
clusters of
galaxies~\cite{Schlickeiser:1987aa,Petrosian:2001ph,Brunetti:2003bh,Brunetti:2007zp},
gamma-ray bursts~\cite{Waxman:1995vg,Dermer:2000gu}, extra-galactic
large-scale jets~\cite{Stawarz:2002uh,Stawarz:2004tq},
blazars~\cite{Katarzynski:2006zc,Giebels:2006ma}, solar
flares~\cite{Petrosian:1999gx,Liu:2004vn,Petrosian:2004ft}, the
interstellar medium~\cite{Simon:1985pz,Seo:1994aa}, the galactic
centre~\cite{Liu:2004zi,Atoyan:2004ix}, supernova
remnants~\cite{Scott:1975aa,Cowsik:1984aa,Fan:2009kr} and even the
recently discovered ``Fermi bubbles''~\cite{Mertsch:2011es} have been
suggested as sites of stochastic acceleration.

The dynamics of the phase-space density $f(\vec{p}, \vec{x}, t)$ of
relativistic particles interacting with a turbulent magnetised plasma
is governed by a Fokker-Planck equation. If the time for pitch-angle
scattering is much smaller than the timescales of interest,
i.e. acceleration, escape and loss times, it suffices to consider the
isotropic part of the phase-space density, $f(p, \vec{x}, t)$ where $p
= \sqrt{ \vec{p}^2}$. Here we consider only relativistic energies such
that energy $E$ and momentum $p$ are essentially the same, $E = p c$.
Furthermore, we constrain ourselves to a distribution function $f(p,
t)$ independent of position, e.g. the spatial average if the rates
only change slowly over the acceleration region. In the special case
of transport coefficients constant in time (see discussion in
Sec.~\ref{sec:GreensFunction}), this constraint can be relaxed as the
spatial and the momentum parts of the problem decouple, see
e.g. Ch.~14 of Ref.~\cite{Schlickeiser:2002pg}.

Stochastic acceleration is a biased diffusion in momentum space and
results in a broadening and systematic shift of the injection spectrum
to higher momenta. In quasi-linear theory, particles interact
resonantly with turbulent plasma waves of a range of wave-lengths
similar to the particles gyroradius $r_g$, and the resulting momentum
diffusion coefficient $D_{pp}$ depends on the spectrum of
turbulence. In particular, if we consider resonant interaction with a
power law spectrum $\mathcal{W}(k) \propto k^{-q}$ of MHD waves of
velocity $v_{\rm A} = \beta_{\rm A} c$, the diffusion coefficient
takes the
form~\cite{Melrose:1968aa,Kulsrud:1969zz,Schlickeiser:1989aa}
\begin{equation}
D_{pp} = \frac{\zeta \beta_{\rm A}^2 p^2 c}{r_g^{2-q} \lambda_2^{q-1}} \, , \label{eqn:Dpp}
\end{equation}
where $\zeta = (\delta B)^2 / B^2$ is the energy in turbulence
$(\delta B)^2 = \int_{k_1}^{k_2} \dd k' \mathcal{W} (k')$, compared to
the energy of the background magnetic field $B$ and $\lambda_2 = 2 \pi
/ k_1$ is the longest wave-length of the MHD modes. The momentum
dependence of the diffusion coefficient reflects the power law
behaviour of the turbulence spectrum, $D_{pp} \propto p^q$. In the
Kolmogorov and Kraichnan phenomenologies the spectral index is $q =
5/3$ and $3/2$, respectively~\cite{Zhou:1990aa}. A well known and
particularly straightforward approximation is the so-called
``hard-sphere'' limit in which $D_{pp} \propto p^2$. If one however
assumes the scattering mean-free path to be equal to the gyroradius
one finds $D_{pp} \propto p$ (Bohm limit).

In eq.~\ref{eqn:Dpp} we have omitted a numerical factor which depends
on $q$ as well as on the magnetic and cross helicity, e.g. in the case
of slab-Alfv\'en turbulence~\cite{Dung:1990bb}. For \mbox{$1.5
  \lesssim q \lesssim 2.5$} and the simplest case of
right-/left-handed waves propagating parallel and antiparallel with
the same power, this factor is $\mathcal{O}(1)$. If Alfv\'en waves
however only propagate into one direction, the momentum diffusion
coefficient vanishes.  Furthermore, also allowing for obliquely
propagating modes one needs to consider that Alfv\'enic turbulence is
inherently anisotropic~\cite{Goldreich:1995zz,Cho:2002qi} which
reduces the efficiency of second-order acceleration. In contrast,
fast-mode waves or turbulence at super-Alfv\'enic scales (which
cascades hydrodynamically) are isotropic~\cite{Cho:2002qi} and probably
dominate 2nd-order Fermi
acceleration~\cite{Schlickeiser:1998zz,Cho:2005mb}.

Despite the relevance of stochastic acceleration for a variety of
astrophysical environments, however, only solutions for a somewhat
limited range of $q$ and assumptions about the acceleration, escape
and loss rates have been presented. The first solution of the
Fokker-Planck equation including Bremsstrahlung losses but only a
constant escape rate was given for the hard-sphere limit ($q =
2$)~\cite{Kaplan:1955aa} using Mellin transforms with respect to
particle energy. Still for $q=2$, this was later
extended~\cite{Kardashev:1962aa} to include time-dependent escape but
no Bremsstrahlung. A comprehensive review of known time-dependent
solutions (and a discussion of the importance of boundary conditions
and its relevance for the steady state spectrum) was presented in
Ref.~\cite{Park:1995aa}, although only for time-independent rates and
limited values of $q$. Finally, the transport equation has been solved
for arbitrary $q$~\cite{Schlickeiser:2002pg} and extensively
discussed~\cite{Becker:2006nz,Fedorov:2010zz}), however, again only
for time-independent rates. We conclude that for \emph{time-dependent}
rates, a solution only exists for $q=2$ whereas for general $q$ the
loss and gain rates must be assumed to be \emph{time-independent}.

For certain classes of environments, however, this assumption is
difficult to justify. For example for very young ($\lesssim 100 \,
\text{yr}$) supernova remnants, the acceleration, escape and adiabatic
loss rates are expected to vary on timescales less than or equal to
the acceleration or loss
times~\cite{Gull:1973aa,Cowsik:1984aa}. Another example are blazar
jets with variability on timescales down to
minutes~\cite{Tammi:2008vg}.  What is therefore needed is the
time-dependent solution of the Fokker-Planck equation for arbitrary
$q$ and allowing for \emph{time-dependent} energy gain and loss rates.
In this paper, we present a time-dependent solution of the
Fokker-Planck equation for general $q$ considering
\emph{time-dependent} stochastic acceleration, escape and adiabatic
losses or gains. We employ a combination of integral transforms to
reduce the transport equation to the heat equation in
\mbox{$(\gamma+1)$-dimensional} spherical coordinates (where $\gamma$
is a function of $q$). Our result improves beyond solutions known to
date and constitutes an important contribution to particle transport
theory.

In Section~\ref{sec:GreensFunction} of this paper we derive the
Green's function of the Fokker-Planck equation with time-dependent
stochastic acceleration, diffusive or convective escape and adiabatic
loss/gain terms for arbitrary $q$, using a combination of integral
transforms. We apply this newly found solution to four specific (toy)
models of time-dependencies in Section~\ref{sec:Examples}. In
particular, assuming all rates to be constant the solution of
Ref.~\cite{Becker:2006nz} is recovered which constitutes a non-trivial
test of our calculation. We conclude in Section~\ref{sec:Conclusion}
with some remarks on boundary conditions and the existence of the
steady state solution.

\section{The Green's function}
\label{sec:GreensFunction}

We start from the transport equation for the isotropic and spatially
averaged phase space density $f(p,t)$ in flux conservation
form~\cite{Schlickeiser:1989aa,Schlickeiser:1989bb},
\begin{equation}
\frac{\partial f(p,t)}{\partial t} = - \frac{1}{p^2} \frac{\partial}{\partial p} \left( p^2 \left( - D_{pp} (p,t) \frac{\partial f(p,t)}{\partial p} + A(p,t) f (p,t) \right) \right) - \frac{f(p,t)}{\tau (p,t) } + \frac{S(p,t)}{4 \pi p^2} \, . \label{eqn:transport1}
\end{equation}
The terms on the right hand side describe biased diffusion in momentum
space with a diffusion coefficient $D_{pp} (p,t)$, additional energy
gain/loss processes with a rate $A(p,t)$, global escape with a rate
$1/\tau(p,t)$ and injection with a rate $S(p,t)/(4 \pi p^2)$.

For adiabatic losses/gains the rate is proportional to $p$, so we
define $a(t)$ by
\begin{equation}
A(p,t) = m c \left( \frac{p}{m c} \right) a(t) \, ,
\end{equation}
where $m$ is the mass of the particle. We note that this form can in
principle also account for bremsstrahlung losses and gains by
1st-order Fermi acceleration at shocks. However, with this particular
momentum dependence, cooling by synchrotron radiation or inverse
Compton scattering cannot be accounted for because the momentum
dependence of the loss rate $A(p,t)$ is more complex then the usually
assumed $p^2$ which is only valid in the Thomson regime. While in
certain limits, e.g. in the steady-state case (see
e.g.~\cite{Schlickeiser:2002pg}), analytical solutions might be
possible, the fully general case is only amenable to numerical
approaches. If the diffusion in momentum space is due to resonant
interactions with MHD waves, the momentum dependence of the diffusion
coefficient $D_{pp}$ reflects the spectrum of the turbulence
cascade. In particular, assuming the spectral energy density $W(k)$ to
be $\propto k^{-q}$, we have $D_{pp} \propto p^q$, and we define the
acceleration rate $k(t)$ by
\begin{equation}
D_{pp} (p,t) = k(t) (m c)^2 \left( \frac{p}{m c} \right)^q \, ,
\end{equation}
where q = 1 for Bohm diffusion, $q = 3/2$ for Kraichnan turbulence, $q
= 5/3$ for Kolmogorov turbulence and $q = 2$ in the hard-sphere
approximation.

\subsection{Diffusive escape}

Assuming that the interactions with the same turbulent MHD waves
dominate spatial diffusion and therefore the diffusive escape from the
acceleration region, fixes the energy dependence of the escape
rate. In particular, the time for diffusive escape from a region of
spatial extent $L$ is $\tau \sim L^2/D_{xx}$ where the spatial
diffusion coefficient $D_{xx}$ is related to the momentum diffusion
coefficient $D_{pp}$ by $D_{xx} D_{pp} = \xi v_{\text{A}}^2 p^2$ with
$v_{\text{A}}$ the Alfv\'en velocity. Here, $\xi$ is a factor that
depends on $q$ as well as the magnetic and cross helicity of the
magnetic turbulence. In this case, $\tau \propto p^{q-2}$ and we
define $\tau_{\text{d}} (t)$ through the relation
\begin{equation}
\tau (p,t) = \tau_{\text{d}} (t) \left( \frac{p}{m c} \right)^{q-2} \, .
\end{equation}

We are now looking for the Green's function to the transport
equation~\ref{eqn:transport1}, that is $f(p,t)$ for mono-energetic,
impulsive injection \mbox{$S(p,t) = \delta ( p - p_0) \delta (t-t_0) /
  (4 \pi p^2)$}. Introducing the dimensionless momentum variable $x
\equiv p / (m c)$, the transport equation reads,
\begin{align}
& \frac{\partial f}{\partial t} + 3 \, a(t) f + \left( a(t) - (2+q) k(t) x^{q-2} x \right) x \frac{\partial f}{\partial x} - k(t) x^q \frac{\partial^2 f}{\partial x^2} \nonumber \\
 &+ \frac{f}{\tau_{\text{d}}(t)} x^{2-q}  = \frac{ \delta ( x - x_0) \delta (t - t_0) }{(m c)^3 4 \pi x_0^2} \, , \label{eqn:transport2}
\end{align}
with $x_0 \equiv p_0 / (m c)$. 

We make the substitutions,
\begin{align}
\rho(x,t) &= 2 x^{(2-q)/2} \sqrt{g(t) \psi(t)} \quad \text{where} \quad g(t) = \exp{ \left[ - (2-q) \int_{t_0}^t \dd t' a(t') \right] } \, , \\
\eta &= \varphi(t) \, \quad \text{and} \quad f = \hat{f} \exp{ \left[ y \, \alpha(t) - \int_{t_0}^t \dd t' \lambda(t') \alpha(t') - 3 \int_{t_0}^t \dd t' a(t') \right] } \, ,
\end{align}
where we choose
\begin{align}
\frac{\dd \alpha}{\dd t} 	&= (2-q)^2 \alpha(t)^2 k(t) g(t) - \frac{1}{\tau_{\text{d}}(t) g(t)} \, , & \label{eqn:Riccati} \\
\psi(t) &= \exp{ \left[ 2 (2-q)^2 \int_{t_0}^t \dd t' \alpha(t') k(t') g(t') \right] } \label{eqn:psi} \, , \\
\varphi(t) &= (2-q)^2 \int_{t_0}^t \dd t' k(t') g(t') \psi(t') \label{eqn:phi} \, .
\end{align}
Eq.~\ref{eqn:transport2} then transforms to
\begin{equation}
\frac{\partial \hat{f}}{\partial \eta} =
\frac{\partial^2 \hat{f}}{\partial \rho^2} + \gamma \frac{1}{\rho} \frac{\partial \hat{f}}{\partial \rho}  + \exp{ \left[ - x \alpha + \int_{t_0}^t \dd t' \lambda \alpha +3 \int_{t_0}^t \dd t' a(t') \right] } (\varphi(t))^{-1} \delta(x-x_0) \delta(t-t_0) \, , \label{eqn:transformed}
\end{equation}
that is the heat equation with spherical symmetry in $(\gamma +
1)$-dimensional spherical coordinates where $\gamma =
(4+q)/(2-q)$. Equation~\ref{eqn:Riccati} is a special case of the
Riccati equation and solutions $\alpha(t)$ for explicit $k(t), g(t)$
and $\tau_{\text{d}} (t)$ are known and have been compiled in
Refs.~\cite{Murphy:1960aa,Kamke:1977aa, Polyanin:1995aa}.

The bounded Green's function, i.e. the solution to
\begin{equation}
\frac{\partial \hat{f}}{\partial y} =
\frac{\partial^2 \hat{f}}{\partial \rho^2} + \gamma \frac{1}{\rho} \frac{\partial \hat{f}}{\partial \rho}  + \delta(\rho-\rho_0) \delta(\eta-\eta_0)
\end{equation}
that remains finite for all $\rho$ and $\eta > 0$, is
\begin{equation}
\hat{f}(\rho, \rho_0, \eta, \eta_0) = \frac{\rho_0}{2 (\eta - \eta_0)} \exp{ \left[ - \frac{ \rho^2 + \rho_0^2 }{4 (\eta - \eta_0) } \right] } I_{\frac{\gamma - 1}{2}} \left( \frac{ \rho \rho_0 }{2 (\eta - \eta_0) } \right) \left( \frac{\rho}{\rho_0} \right)^{(1-\gamma)/2} \, ,
\end{equation}
with $ I_{(\gamma-1)/2}$ the modified Bessel function of the first
kind. Resubstituting for $\rho = \rho(x,t)$ and $\eta = \eta(t)$ one
finds for $q \not = 2$,
\begin{align}
f &= \frac{1}{(m c)^3 4 \pi x_0^2} \frac{2-q}{x_0} \exp{ \left[ - \frac{3}{2} \int_{t_0}^t \dd t' a(t') \right] } \exp{ \left[ x^{2-q} g \alpha - x_0^{2-q} \alpha_0 \right] } \frac{\left( x x_0 \right)^{\frac{2-q}{2}} \sqrt{ g \psi }}{ \varphi } \nonumber \\
& \times \exp{ \left[ - \frac{x^{2-q} g \psi + x_0^{2-q} \psi_0}{\varphi} \right] } I_{\frac{1+q}{2-q}} \left[ \frac{2 \left( x x_0 \right)^{\frac{2-q}{2}} \sqrt{ g \psi }}{ \varphi } \right] \left( \frac{x}{x_0} \right)^{-3/2} \, . \label{eqn:GreensFunctionQnotEqual2}
\end{align}
which, together with eqs.~\ref{eqn:Riccati} - \ref{eqn:phi},
constitutes the main result of this paper.

For the hard-sphere approximation, one needs to carefully take the
limit $q \rightarrow 2$. After some tedious algebra including a number
of non-trivial cancellations, one arrives at the known
result~\cite{Kardashev:1962aa,Cowsik:1984aa},
\begin{equation}
f = \frac{1}{(m c)^3 4 \pi x_0^2} \frac{1}{x} \frac{1}{\sqrt{4 \pi}} \frac{ \exp{ \left[ - \int_{t_0}^t \frac{\dd t'}{\tau_{\text{d}}} \right] } }{ \sqrt{ \int_{t_0}^t \dd t' k } } \exp { \left[ - \frac{ \left( \ln x -\ln x_0 - \int_{t_0}^t \dd t' a - 3 \int_{t_0}^t \dd t' k \right)^2}{4 \int_{t_0}^t \dd t' k } \right]  } . \label{eqn:fQequal2}
\end{equation}
We stress that this result has been derived in a way completely
independent from the one in
Refs.~\cite{Kardashev:1962aa,Cowsik:1984aa} and therefore constitutes
a valuable test of our calculation.

\subsection{Convective escape}

If we assume an energy-independent escape time as is for example the
case if convection out of the acceleration zone is dominating over
diffusive escape,
\begin{equation}
\tau_{\text{esc}}(p,t) \equiv \tau_{\text{c}} (t) \, ,
\end{equation}
the Green's function takes a form similar to
eq.~\ref{eqn:GreensFunctionQnotEqual2},
\begin{align}
f &= \frac{1}{(m c)^3 4 \pi x_0^2} \frac{2-q}{x_0} \exp{ \left[ - \int_{t_0}^t \frac{\dd t'}{\tau_{\text{c}}} \right] } \exp{ \left[ - \frac{3}{2} \int_{t_0}^t \dd t' a(t') \right] } \frac{\left( x x_0 \right)^{\frac{2-q}{2}} \sqrt{ g }}{ (q-2)^2 \int_{t_0}^t \dd t' k g } \nonumber \\
& \times \exp{ \left[ - \frac{x^{2-q} g + x_0^{2-q}}{(q-2)^2 \int_{t_0}^t \dd t' k g} \right] } I_{\frac{1+q}{2-q}} \left[ \frac{2 \left( x x_0 \right)^{\frac{2-q}{2}} \sqrt{ g }}{ (q-2)^2 \int_{t_0}^t \dd t' k g } \right] \left( \frac{x}{x_0} \right)^{-3/2} \, . \label{eqn:GreensFunctionQnotEqual2ConvecEscape}
\end{align}

For $q=2$, the result is again eq.~\ref{eqn:fQequal2} since in the
hard-sphere approximation the escape time for diffusive and convective
escape are both energy-independent, $\tau_{\text{c}} \sim
\tau_{\text{d}}$.

\section{Examples}
\label{sec:Examples}

\subsection{Constant acceleration and escape rates, but no adiabatic losses}

We assume that the acceleration and escape rates are constant, $k(t)
\equiv k_0$, $\tau_{\text{d}}(t) \equiv \tau_{\text{d}0}$, and that
there are no adiabatic losses, $a(t) \equiv 0$. The Riccati
equation~\ref{eqn:Riccati} is solved by
\begin{equation}
\alpha = - \frac{1}{\varepsilon} \frac{1}{\sqrt{k_0 \tau_{\text{d0}}}} \, ,
\end{equation}
and we find for $\psi(t)$ and $\varphi(t)$
\begin{align}
\psi(t) 		&= \exp{ \left[ -2 (2-q) \sqrt{\frac{k_0}{\tau_{\text{d0}}}} (t-t_0) \right] } \, , \\
\varphi(t) 	&= \frac{1}{2} (2-q) \sqrt{k_0 \tau_{\text{d0}}} (1-\psi) \, .
\end{align}
For $q \not = 2$, the particle density $n (x,x_0,t,t_0) = 4 \pi x^2 f
(x,x_0,t,t_0)$ for impulsive injection reads
\begin{align}
n (x,x_0,t,t_0) =& \frac{2-q}{x_0} \sqrt{ \frac{x}{x_0} } \frac{ 2 (x x_0)^{(2-q)/2} \sqrt{\psi} }{ (2-q) \sqrt{k_0 \tau_{\text{d0}}} (1-\psi) } \exp{ \left[ - \frac{ (x^{2-q} + x_0^{2-q}) (1+\psi)  }{(2-q) \sqrt{k_0 \tau_{\text{d0}}} (1-\psi)} \right] } \nonumber \\
& \times I_{\frac{1+q}{2-q}} \left[ \frac{ 4 (x x_0)^{(2-q)/2} \sqrt{\psi} }{ (2-q) \sqrt{k_0 \tau_{\text{d0}}} (1-\psi) }  \right]  \, ,\label{eqn:nConstQnot2}
\end{align}
and for $q = 2$ this reduces to
\begin{equation}
n (x,x_0,t,t_0) = \frac{1}{p} \frac{1}{\sqrt{4 \pi}} \frac{ \exp{ \left[ - \frac{t-t_0}{\tau_{\text{d}0}} \right] } }{ \sqrt{ k_0 (t-t_0) } } \exp { \left[ - \frac{ \left( \ln x - \ln x_0 - 3 k_0 (t-t_0) \right)^2 }{4 k_0 (t-t_0) } \right]  } \, . \label{eqn:nConstQequal2}
\end{equation}
Equations~\ref{eqn:nConstQnot2} and~\ref{eqn:nConstQequal2} are
identical to eqs.~46 and 49 of Ref.~\cite{Becker:2006nz} (setting
their $a \equiv 0$), respectively (see also
\cite{Schlickeiser:2002pg}). Their result was however derived assuming
constant acceleration rate and escape time. Reproducing this result as
a special case of our more general solution therefore constitutes a
non-trivial test of our calculation.

\begin{figure}
\includegraphics[width=\textwidth]{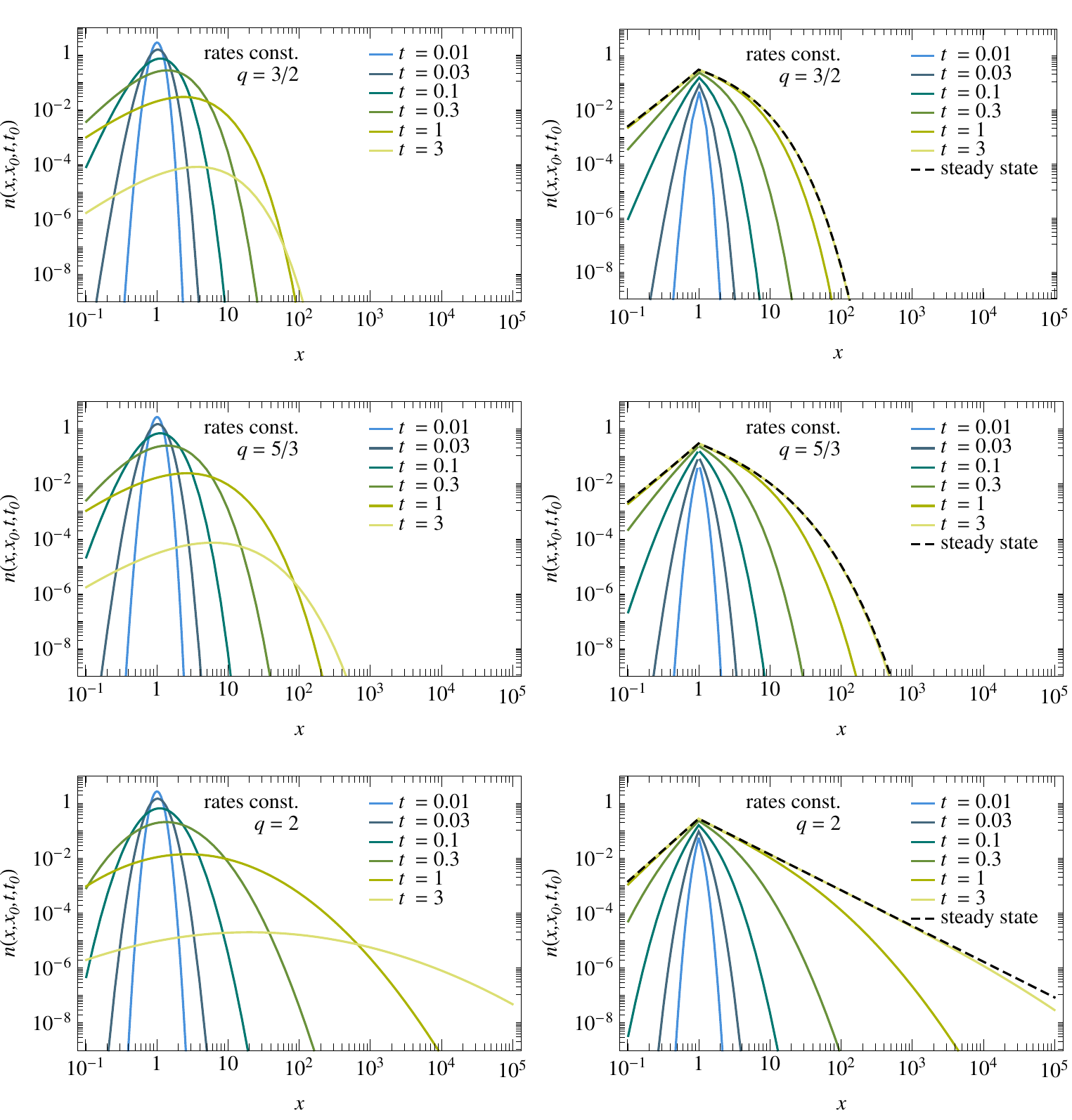}
\caption{Particle spectrum $n (x,1,t,0)$ for impulsive (left panel)
  and steady injection (right panel) for Kraichnan turbulence (first
  line), Kolmogorov turbulence (second line) and the hard-sphere
  approximation (last line) and assuming $a=0$ and $\tau_{\text{d0}} =
  1$. The solid lines are for fixed times \mbox{$t = 0.01, 0.03, 0.1,
    0.3, 1, 3$} and the dashed lines denote the steady state
  spectrum. All timescales (rates) are in units of $1/k_0$ ($k_0$).}
\label{fig1}
\end{figure}

In the left column of Fig.~\ref{fig1} (compare with Fig.~2 of
Ref.~\cite{Becker:2006nz}) we show the particle spectrum for impulsive
injection, $n (x,1,t,0)$, for Kraichnan turbulence, Kolmogorov
turbulence and the hard-sphere approximation. We fixed
$\tau_{\text{d0}} = 1$ and all timescales (rates) are in units of
$1/k_0$ ($k_0$). Diffusion and advection in momentum space lead to a
broadening of the spectrum and monotonous increase of the mean energy
with time. Since in the hard-sphere approximation the acceleration
timescale $p^2/D_{pp}$ is the same for all momenta, the spectrum (in
the logarithmic momentum variable, $\log x$) is even with respect to
the mean logarithmic energy, $(\log x_0 + 3 k_0 (t-t_0))$. For
Kraichnan and Kolmogorov turbulence the behaviour is qualitatively
different as the acceleration time is now increasing with energy. At
higher energies, the spectrum becomes gradually softer and
asymmetric. For a fixed escape time, the spectrum rolls over at much
smaller energies than in the hard-sphere case.

In the right column of Fig.~\ref{fig1} we show the spectra for steady
injection,
\begin{equation}
n (x,x_0,t,t_0) = \int_{t_0}^t \dd t' f (x,x_0,t,t') \, ,
\end{equation}
including the steady state spectrum, i.e. the limit $t \rightarrow
\infty$. In general, it is not clear whether this integral in fact
converges for $t \rightarrow \infty$ but for the case of constant
rates, injection and escape exactly balance each other. We note that
for the hard-sphere approximation, acceleration and escape rates have
the same momentum behaviour and consequently there is no preferred
momentum scale. This leads to a power law steady state spectrum
whereas for Kraichnan and Kolmogorov phenomenology the spectrum
exhibits a long exponential roll-over with a characteristic momentum
defined by the equality of acceleration and escape times.

\subsection{Constant acceleration rate}

We use the relation between the space and momentum diffusion
coefficients, $D_{xx} D_{pp} = \xi v_\text{A}^2 p^2$, to express the
escape time $\tau(t,p) = \tau_{\text{d}}(t) x^{q-2}$ in terms of the
acceleration rate $k(t)$,
\begin{equation}
\frac{1}{\tau_{\text{d}}(t)} = \frac{\xi v_A^2 (t)}{k(t) L^2} = \frac{1}{k(t)} \frac{\xi}{\rho_m} \left( \frac{B(t)}{L(t)} \right)^2 \, ,
\end{equation}
where $L(t)$ is the size of the acceleration region and the Alfv\'en
velocity \mbox{$v_A (t) = B(t) / \sqrt{\rho_m}$} with $B$ the
background magnetic field and $\rho_m$ the thermal gas mass
density. Specifying the adiabatic loss/gain rate, $a(t)$, to the case
of an expanding/contracting flux tube~\cite{Cowsik:1984aa},
\begin{equation}
a(t) = \frac{1}{3} \left( \frac{\dd \ln L(t)}{\dd t} - \frac{\dd \ln B(t)}{\dd t} \right) \quad \Rightarrow \quad g = \exp{ \left[ -3 \int_{t_0}^t \dd t' a(t') \right] } = \frac{L_0}{L(t)} \frac{B(t)}{L_0} \, ,
\end{equation}
with $L_0 = L(t_0)$ and $B_0 = B(t_0)$, we can write
\begin{equation}
\frac{1}{\tau_{\text{d}}(t)} = \frac{1}{k(t)} \frac{\xi}{\rho_m} \left( \frac{B_0}{L_0} \right)^2 g^2(t) \, .
\end{equation}
The Riccati equation~\ref{eqn:Riccati} now reads,
\begin{equation}
\frac{\dd \alpha}{\dd t} = \alpha^2 (q-2)^2 k g - \frac{\xi}{\rho_m} \left( \frac{B_0}{L_0} \right)^2 \frac{g}{k} \, . 
\label{eqn:Riccati+quasilinear}
\end{equation}
For a constant acceleration rate, $k(t) \equiv k_0$, this is solved by
\begin{equation}
\alpha(t) \equiv \alpha_0 = \frac{1}{q-2} \sqrt{\frac{\xi}{\rho_m}} \frac{B_0}{L_0} \frac{1}{k_0} \, ,
\end{equation}
leading to
\begin{align}
\psi(t) 		&= \exp{ \left[ 2 (2-q)^2 \alpha_0 k_0 \int_{t_0}^t \dd t' g(t') \right] } \, , \\
\varphi(t) 	&= \frac{1}{2 \alpha_0} (\psi - 1) \, .
\end{align}

\begin{figure}
\includegraphics[width=\textwidth]{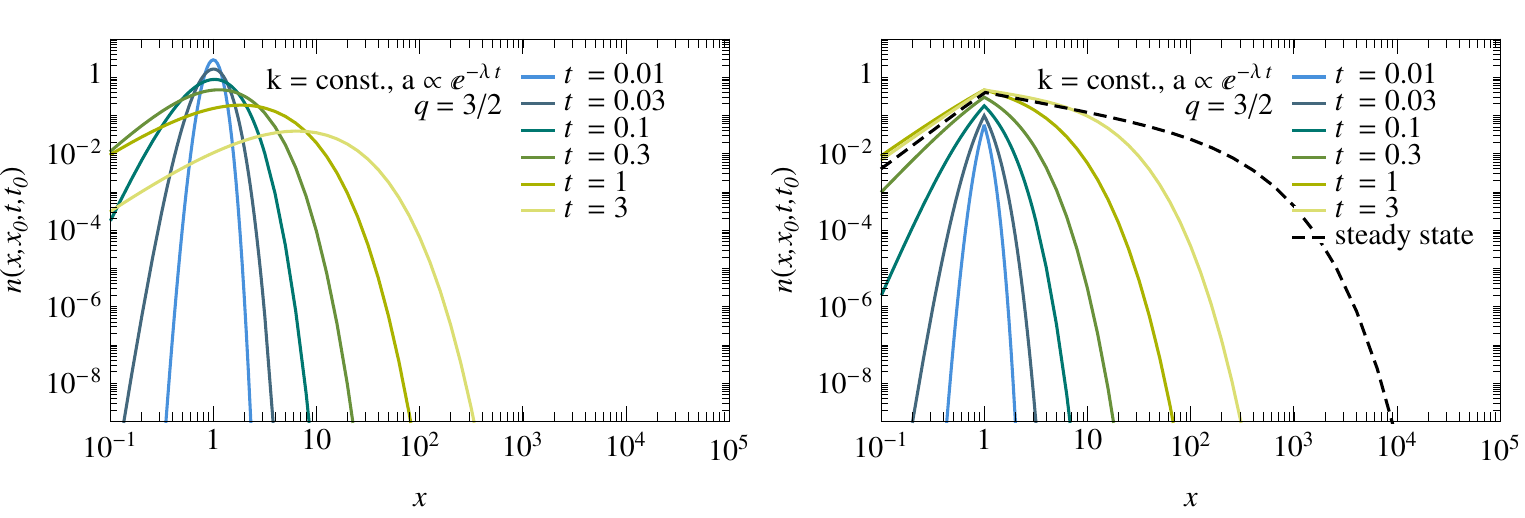}
\caption{Particle spectrum $n (x,1,t,0)$ for impulsive (left panel)
  and steady injection (right panel) for Kraichnan turbulence and
  assuming $a=a_0 \exp{ \left[ - \lambda t \right] }$ with $a_0 = -1$,
  $\lambda = 0.5$ and $( \xi B_0^2 / L_0^2 \rho_m ) = 0.1$. The solid
  lines are for fixed times \mbox{$t = 0.01, 0.03, 0.1, 0.3, 1, 3$}
  and the dashed line denotes the steady state spectrum. All
  timescales (rates) are in units of $1/k_0$ ($k_0$).}
\label{fig2}
\end{figure}

In Fig.~\ref{fig2} we show the particle spectrum $n (x,1,t,0)$ for
impulsive and steady injection and Kraichnan turbulence. We have
chosen $L(t)$ and $B(t)$ such that
\begin{equation}
a(t) = a_0 \exp{ \left[ - \lambda t \right] } \, ,
\end{equation}
with $a_0 = -1$, $\lambda = 0.5$ and $( \xi B_0^2 / L_0^2 \rho_m ) =
0.1$. Again, all times (rates) are understood to be in units of
$1/k_0$ ($k_0$). For early times ($t \ll 1$), the acceleration and
adiabatic loss rate are similar, $| a(t) | \approx k_0$ and escape can
be neglected. For intermediate times ($t \sim 1$), the loss rate
starts declining which leads to a rather hard spectrum. Finally, for
late times ($t \gg 1$), escape becomes important such that the
spectrum does not extend to any higher energies. The late onset of
escape also reflects in very hard spectra for steady injection and a
steady state spectrum extending over several orders of magnitude.

\subsection{Constant adiabatic loss rate and exponentially decreasing acceleration rate}

We assume that the adiabatic loss rate $a(t) < 0$ is constant, $a(t)
\equiv a_0$. This is a fair assumption for environments with a blast
wave where the shock radius and speed are \mbox{$r(t) \propto
  t^{2/5}$} and $v(t) \propto t^{-3/5}$ such that the adiabatic loss
rate $a(t) \propto (r^2 v) \propto t^{1/5}$ has a weak time
dependence. We find for $g(t)$,
\begin{equation}
g = g_0 \exp{ \left[ \lambda t \right] } \quad \text{with} \quad \lambda = - (2-q) a_0 >0 \quad \text{and} \quad g_0  = - \lambda t_0 \, .
\end{equation}

We assume further that $k(t)$ is of the form $k_0 \exp { \left[ -
    \kappa t \right] }$ and that $\kappa = \lambda$. (In fact, this
can be easily extended to $\kappa \not = \lambda$ but for clarity we
here constrain ourselves to the case $\kappa = \lambda$.) The Riccati
equation~\ref{eqn:Riccati+quasilinear} now reads
\begin{equation}
\frac{\dd \alpha}{\dd t} = \alpha^2 (q-2)^2 k_0 g_0 - \frac{\xi}{\rho_m} \left( \frac{B_0}{L_0} \right)^2 \frac{g_0}{k_0} \exp { \left[ 2 \lambda t \right] }\, . 
\label{eqn:Riccati+quasilinear2}
\end{equation}
and substituting $y = \exp { \left[ 2 \lambda t \right] } $, $w =
-1/\alpha$ leads to the standard form
\begin{equation}
\frac{\dd w}{\dd y} = \mathcal{A}^2 w^2 + \mathcal{B} \frac{1}{y} \quad \text{with} \quad \mathcal{A} \equiv - \frac{1}{\rho_m} \left( \frac{B_0}{L_0} \right)^2 \frac{g_0}{k_0} \frac{1}{2 \lambda} \quad \text{and} \quad \mathcal{B} \equiv (q-2)^2 \frac{k_0 g_0}{2 \lambda} \, . 
\label{eqn:Riccati+quasilinear3}
\end{equation}
The solution for $w(y)$ is~\cite{Polyanin:1995aa}
\begin{equation}
w = - \frac{1}{\mathcal{A}} \frac{1}{u} \frac{\dd u}{\dd y} \quad \text{where} \quad u = \sqrt{y} J_1 \left( 2 \sqrt{ \mathcal{A} \mathcal{B} } \sqrt{y} \right) \, ,
\end{equation}
with $J_1$ the Bessel function of the first kind. For $\alpha(t)$ and
$\psi(t)$ we find
\begin{equation}
\alpha(t) = \sqrt{\frac{\mathcal{A}}{\mathcal{B}}} \sqrt{y} \frac{J_1 \left( 2 \sqrt{ \mathcal{A} \mathcal{B} } \sqrt{y} \right)}{J_0 \left( 2 \sqrt{ \mathcal{A} \mathcal{B} } \sqrt{y} \right)}
\quad \text{and} \quad 
\psi(t) = \left( \frac{J_1 \left( 2 \sqrt{ \mathcal{A} \mathcal{B} } \sqrt{y_0} \right)}{J_1 \left( 2 \sqrt{ \mathcal{A} \mathcal{B} } \sqrt{y} \right)} \right)^2 \, ,
\end{equation}
where $y_0 = y(t_0)$. $\varphi(t)$ is again given by eq.~\ref{eqn:phi}.

\begin{figure}
\includegraphics[width=\textwidth]{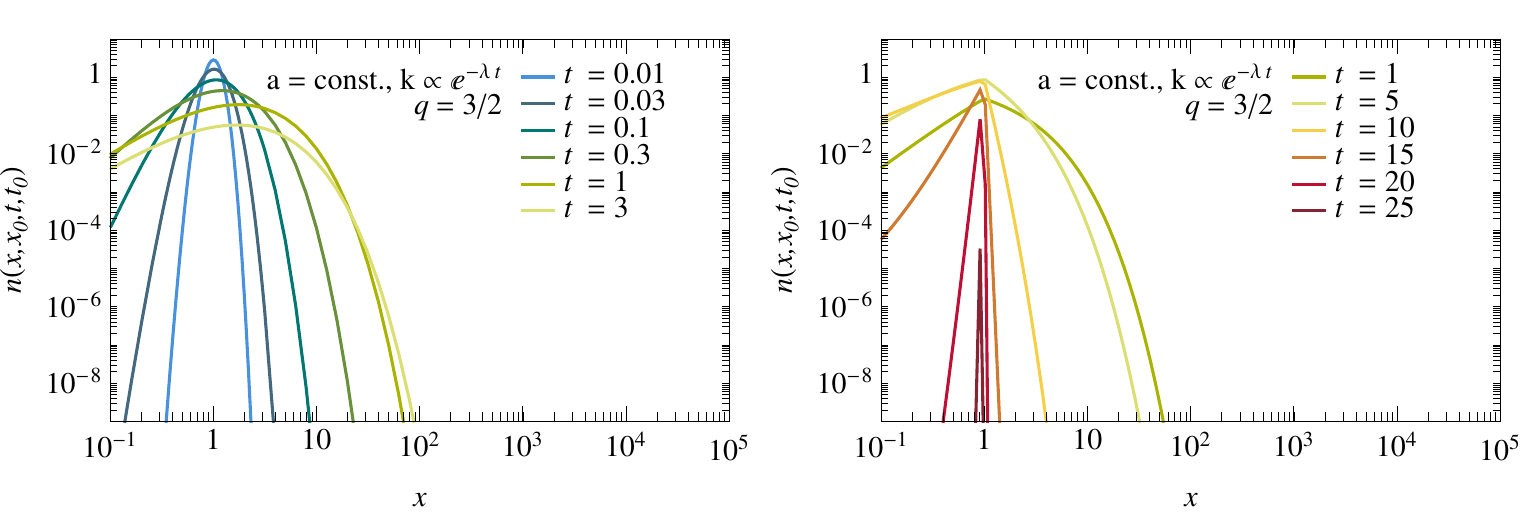}
\caption{Particle spectrum $n (x,1,t,0)$ for impulsive (left panel)
  and steady injection (right panel) for Kraichnan turbulence. We
  assumed $a = a_0 = \text{const.}$ and $k=k_0 \exp{ \left[ - \lambda
      t \right] }$ with $a_0 = -0.6$ and $( \xi B_0^2 / L_0^2 \rho_m ) =
  0.2$. The solid lines are for fixed times \mbox{$t = 0.01, 0.03,
    0.1, 0.3, 1, 3$}. All timescales (rates) are in units of $1/k_0$
  ($k_0$).  }
\label{fig3}
\end{figure}

In the left panel of Fig.~\ref{fig3} we show the particle spectrum $n
(x,1,t,0)$ for impulsive injection for Kraichnan turbulence and for
$\lambda = 0.3$ and $( \xi B_0^2 / L_0^2 \rho_m ) = 0.2$. All
timescales (rates) are in units of $1/k_0$ ($k_0$). For early times
($t \lesssim 0.1$), the Green's function behaves similarly as in the
cases considered before. For intermediate times ($0.1 \lesssim t
\lesssim 1$), however, the acceleration rate has started decreasing
while the adiabatic loss rate stays constant. For even later times,
escape becomes dominant since its rate is $\propto g^2 \propto \exp{
  \left[ 2 \lambda t \right] }$ and the Green's function becomes
heavily suppressed. This behaviour is even more prominent in the
spectrum for steady injection, see right panel of Fig.~\ref{fig3}. We
show the particle spectrum for intermediate and late times ($t \gtrsim
1$) only since for early times ($t \lesssim 1$) it looks very similar
to the previously considered examples. Around $t \approx 5$ the
spectrum starts becoming noticeably asymmetric because of the
dominance of adiabatic losses over acceleration. The escape rate
$1/\tau_{\text{d}} \propto g^2(t) \propto \exp{ \left[ 2 \lambda t
  \right] }$ keeps increasing and particles injected early have
already escaped. However, particles which are injected late ($t
\gtrsim 10$) do not get accelerated anymore and hardly have time to
lose energy before escaping. The resulting spectrum therefore
converges against the $\delta$-like injection though with decreasing
amplitude.

\subsection{No escape}
\label{sec:NoEscape}

Assuming a vanishing escape rate, $\tau_{\text{d}} \rightarrow
\infty$, the Ricatti equation can be directly integrated,
\begin{align}
\frac{\dd \alpha}{\dd t} = (q-2)^2 \alpha^2 k g \quad \Rightarrow \quad \alpha = \alpha_0 \left(1 - (q-2)^2 \alpha_0 \int_{t_0}^t \dd t' k g \right)^{-1} \, .
\end{align}
Furthermore, $\psi(t)$ and $\varphi(t)$ are
\begin{equation}
\psi(t) =  \left( \frac{\alpha(t)}{\alpha_0} \right)^2
\quad \text{and} \quad 
\varphi(t) = \left( \left( (q-2)^2 \int_{t_0}^t \dd t' k g \right)^{-1} - \alpha_0 \right)^{-1} \, .
\end{equation}

\begin{figure}
\includegraphics[width=\textwidth]{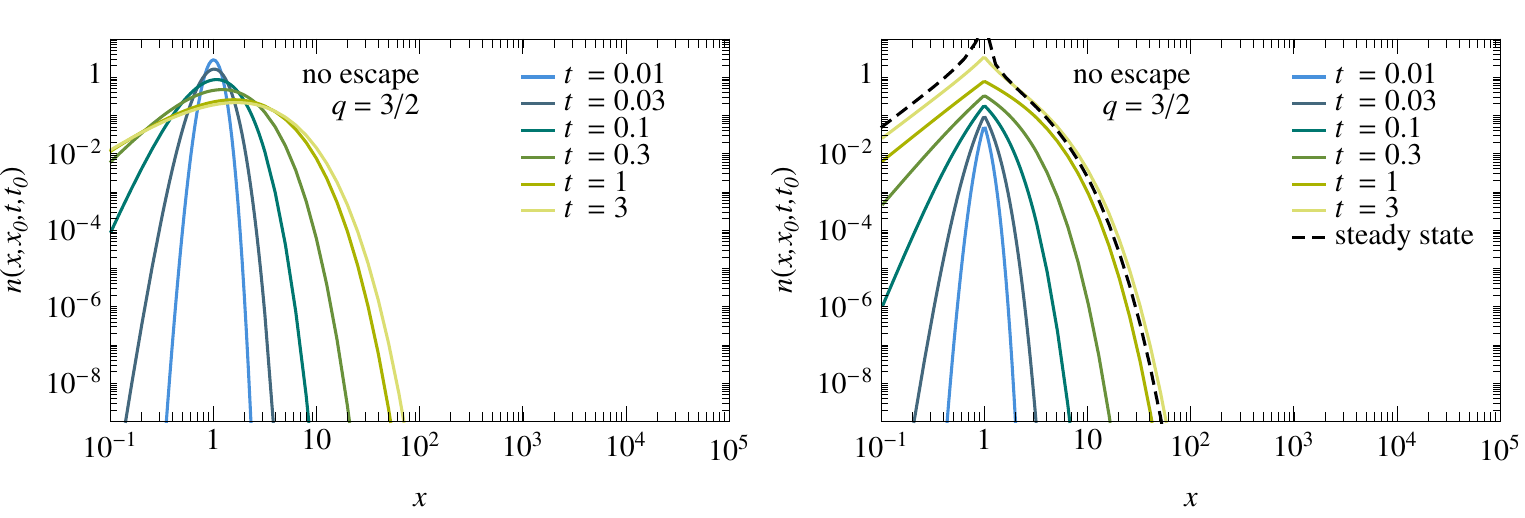}
\caption{Particle spectrum $n (x,1,t,0)$ for impulsive (left panel)
  and steady injection (right panel) for Kraichnan turbulence. We
  assumed $\tau_{\text{d}} \rightarrow \infty$, $a(t) = a_0 \exp{
    \left[ - t / \lambda \right] }$ and $k(t) = k_0 \exp{ \left[ - t /
      \kappa \right] }$ with $a_0 = -0.5$, $\lambda = 2$, $\kappa =
  1$. The solid lines are for fixed times \mbox{$t = 0.01, 0.03, 0.1,
    0.3, 1, 3$} and the dashed line denotes the steady state
  spectrum. All timescales (rates) are in units of $1/k_0$ ($k_0$).}
\label{fig4}
\end{figure}

For illustration, we consider both $k(t)$ and $a(t)$ to be
exponentially declining, i.e. \mbox{$a(t) = a_0 \exp{ \left[ - t /
      \lambda \right] }$} and $k(t) = k_0 \exp{ \left[ - t / \kappa
  \right] }$. In Fig.~\ref{fig4} we show the particle spectrum $n
(x,x_0,t,t_0)$ for impulsive and steady injection for Kraichnan
turbulence and choose $a_0 = -0.5$, $\lambda = 2$, $\kappa = 1$. All
timescales (rates) are in units of $1/k_0$ ($k_0$). At early times,
the Green's function (Fig.~\ref{fig4}, left panel) again behaves very
similarly to the cases considered above. For late times, however, it
stalls as both $a(t)$ and $k(t)$ approach zero. For momenta $x > 1$
and steady injection, the steady state is reached rather quickly as
only particles injected early enough can reach higher momenta. Close
to the injection momentum however, the spectrum still changes even at
late times due to the injection of new particles. As $t \rightarrow
\infty$, the spectrum reaches its steady state at all momenta except
the injection momentum $x_0$.

\section{Summary and conclusion}
\label{sec:Conclusion}

We have presented a new solution to the Fokker-Planck equation for
stochastic particle acceleration by plasma wave turbulence. In
extension to previously known solutions we allow the rates for
stochastic acceleration, both diffusive and convective escape as well
as adiabatic losses to be time-dependent. Furthermore, we do not need
to constrain ourselves to specific values of the turbulence spectral
index $q$ and can therefore apply this result to the
phenomenologically interesting cases of Kolmogorov ($q = 5/3$) and
Kraichnan ($q = 3/2$) turbulence. We have investigated four examples
to illustrate the qualitatively different behaviour of spectra due to
the time-dependent rates and extended range in $q$. In the first
example we constrained ourselves, however, to constant rates and
neglected adiabatic losses in order to compare our solution to
previously presented results which constitutes a non-trivial test of
our calculation. We have also considered the case of constant
acceleration and exponentially decreasing adiabatic loss rate which
leads to rather hard spectra. In contrast, with exponentially
decreasing acceleration and constant adiabatic losses, the spectra
become very soft and no steady-state solution exists. Finally, for
infinite escape time we have presented an example with exponentially
decreasing acceleration and adiabatic loss rates which leads to rather
soft spectra in particular close to the injection energy.

A few words about boundary conditions are in order. It was pointed
out~\cite{Park:1995aa} that the initial value problem (IVP),
eq.~\ref{eqn:transport1}, with boundary conditions at $x = 0$ and $x
\rightarrow \infty$ is singular (see Ref.~\cite{Park:1995aa} for a
definition and discussion of a singular IVP). Therefore, the status of
the boundary conditions is not clear but must be determined more
carefully. Furthermore, the spectral theory of second order
differential equations proves helpful in investigating the conditions
under which a steady state solution exists. Unfortunately, it is not
possible to analyse the problem at hand in this framework, since the
variable coefficients of the partial differential equation prevent the
formulation of an equivalent boundary-value problem (BVP) of
Sturm-Liouville type. However, the transformed IVP, i.e. the heat
equation type eq.~\ref{eqn:transformed}, must satisfy boundary
conditions for $\hat{f}$ at $\rho \rightarrow 0, \infty$ similar to
those that the original IVP, eq.~\ref{eqn:transport2}, satisfies for
$f$ at $x \rightarrow 0, \infty$ since both are connected by
non-singular transformations. We find that $\rho \rightarrow 0,
\infty$ are both limit points (for a definition see
Ref.~\cite{Park:1995aa}) such that the appropriate boundary conditions
at $\rho \rightarrow 0, \infty$ are simply $|| \hat{f} || < \infty$
which justifies the choice of the bounded solution in deriving
eq.~\ref{eqn:transformed}.  Finally, we note that the discussion about
the existence of the steady state solution cannot be applied here
either due to the variable coefficients. As we have seen in the above
examples, the existence of the steady state is largely determined by
the time-dependence and asymptotic behaviour of the acceleration and
loss rates, $k(t)$, $a(t)$ and $\tau_{\text{d}}(t)$ and therefore
needs to be discussed on a case-by-case basis.

\acknowledgments

The author would like to thank Subir Sarkar, Brian Reville and Alex
Lazarian for helpful discussions and Subir Sarkar for encouragement to
publish this study.

\providecommand{\href}[2]{#2}\begingroup\raggedright\endgroup


\begin{thebibliography}{10}

\bibitem{Fermi:1949ee}
E.~Fermi, {\it {On the Origin of the Cosmic Radiation}},  {\em Phys.Rev.} {\bf
  75} (1949) 1169--1174.

\bibitem{Lacombe:1977aa}
C.~{Lacombe}, {\it {Acceleration of particles and plasma heating by turbulent
  Alfven waves in a radiogalaxy}},  {\em Astron. Astrophys.} {\bf 54} (1977)
  1--16.

\bibitem{Eilek:1979aa}
J.~A. {Eilek}, {\it {Particle reacceleration in radio galaxies}},  {\em
  Astrophys. J.} {\bf 230} (1979) 373--385.

\bibitem{Achterberg:1979aa}
A.~{Achterberg}, {\it {The energy spectrum of electrons accelerated by weak
  magnetohydrodynamic turbulence}},  {\em Astron. Astrophys.} {\bf 76} (1979)
  276--286.

\bibitem{Hardcastle:2008jw}
M.~J. {Hardcastle}, C.~C. {Cheung}, I.~J. {Feain}, and {\L}.~{Stawarz}, {\it
  {High-energy particle acceleration and production of ultra-high-energy cosmic
  rays in the giant lobes of Centaurus A}},  {\em Mont. Not. R. Astron. Soc.}
  {\bf 393} (2009) 1041--1053, [\href{http://xxx.lanl.gov/abs/0808.1593}{{\tt
  arXiv:0808.1593}}].

\bibitem{O'Sullivan:2009sc}
S.~O'Sullivan, B.~Reville, and A.~Taylor, {\it {Stochastic particle
  acceleration in the lobes of giant radio galaxies}},  {\em Mon. Not. R.
  Astron. Soc.} {\bf 400} (2009) 248--257,
  [\href{http://xxx.lanl.gov/abs/0903.1259}{{\tt arXiv:0903.1259}}].

\bibitem{Tramacere:2011qw}
A.~Tramacere, E.~Massaro, and A.~M. Taylor, {\it {Stochastic Acceleration and
  the Evolution of Spectral Distributions in SSC Sources: A Self Consistent
  Modeling of Blazars' Flares}},  {\em Astrophys. J.} {\bf 739} (2011) 66,
  [\href{http://xxx.lanl.gov/abs/1107.1879}{{\tt arXiv:1107.1879}}].

\bibitem{Schlickeiser:1987aa}
R.~{Schlickeiser}, A.~{Sievers}, and H.~{Thiemann}, {\it {The diffuse radio
  emission from the Coma cluster}},  {\em Astron. Astrophys.} {\bf 182} (1987)
  21--35.

\bibitem{Petrosian:2001ph}
V.~{Petrosian}, {\it {On the Nonthermal Emission and Acceleration of Electrons
  in Coma and Other Clusters of Galaxies}},  {\em Astrophys. J.} {\bf 557}
  (2001) 560--572, [\href{http://xxx.lanl.gov/abs/astro-ph/0101145}{{\tt
  astro-ph/0101145}}].

\bibitem{Brunetti:2003bh}
G.~Brunetti, P.~Blasi, R.~Cassano, and S.~Gabici, {\it {Alfvenic reacceleration
  of relativistic particles in galaxy clusters: MHD waves, leptons and
  hadrons}},  {\em Mon. Not. R. Astron. Soc.} {\bf 350} (2004) 1174,
  [\href{http://xxx.lanl.gov/abs/astro-ph/0312482}{{\tt astro-ph/0312482}}].

\bibitem{Brunetti:2007zp}
G.~Brunetti and A.~Lazarian, {\it {Compressible Turbulence in Galaxy Clusters:
  Physics and Stochastic Particle Re-acceleration}},  {\em Mon. Not. R. Astron.
  Soc.} {\bf 378} (2007) 245--275,
  [\href{http://xxx.lanl.gov/abs/astro-ph/0703591}{{\tt astro-ph/0703591}}].

\bibitem{Waxman:1995vg}
E.~Waxman, {\it {Cosmological gamma-ray bursts and the highest energy cosmic
  rays}},  {\em Phys. Rev. Lett.} {\bf 75} (1995) 386--389,
  [\href{http://xxx.lanl.gov/abs/astro-ph/9505082}{{\tt astro-ph/9505082}}].

\bibitem{Dermer:2000gu}
C.~D. Dermer and M.~Humi, {\it {Adiabatic losses and ultra-high energy cosmic
  ray acceleration in gamma ray burst blast waves}},  {\em Astrophys. J.} {\bf
  556} (2001) 479--493, [\href{http://xxx.lanl.gov/abs/astro-ph/0012272}{{\tt
  astro-ph/0012272}}].

\bibitem{Stawarz:2002uh}
L.~Stawarz and M.~Ostrowski, {\it {Radiation from the Relativistic Jet: a Role
  of the Shear Boundary Layer}},  {\em Astrophys. J.} {\bf 578} (2002)
  763--774, [\href{http://xxx.lanl.gov/abs/astro-ph/0203040}{{\tt
  astro-ph/0203040}}].

\bibitem{Stawarz:2004tq}
L.~Stawarz, M.~Sikora, M.~Ostrowski, and M.~C. Begelman, {\it {On
  Multiwavelength Emission of Large-Scale Quasar Jets}},  {\em Astrophys. J.}
  {\bf 608} (2004) 95--107,
  [\href{http://xxx.lanl.gov/abs/astro-ph/0401356}{{\tt astro-ph/0401356}}].

\bibitem{Katarzynski:2006zc}
K.~{Katarzy{\'n}ski}, G.~{Ghisellini}, A.~{Mastichiadis}, F.~{Tavecchio}, and
  L.~{Maraschi}, {\it {Stochastic particle acceleration and synchrotron
  self-Compton radiation in TeV blazars}},  {\em Astron. Astrophys.} {\bf 453}
  (2006) 47--56, [\href{http://xxx.lanl.gov/abs/astro-ph/0603362}{{\tt
  astro-ph/0603362}}].

\bibitem{Giebels:2006ma}
B.~Giebels, G.~Dubus, and B.~Khelifi, {\it {Unveiling the X-ray/TeV engine in
  Mkn 421}},  {\em Astron.Astrophys.} {\bf 462} (2007) 29--41,
  [\href{http://xxx.lanl.gov/abs/astro-ph/0610270}{{\tt astro-ph/0610270}}].

\bibitem{Petrosian:1999gx}
V.~{Petrosian} and T.~Q. {Donaghy}, {\it {On the Spatial Distribution of Hard
  X-Rays from Solar Flare Loops}},  {\em Astrophys. J.} {\bf 527} (1999)
  945--957, [\href{http://xxx.lanl.gov/abs/astro-ph/9907181}{{\tt
  astro-ph/9907181}}].

\bibitem{Liu:2004vn}
S.-M. Liu, V.~Petrosian, and G.~M. Mason, {\it {Stochastic Acceleration of 3He
  and 4He by Parallel Propagating Plasma Waves}},  {\em Astrophys. J.} {\bf
  613} (2004) L81, [\href{http://xxx.lanl.gov/abs/astro-ph/0403007}{{\tt
  astro-ph/0403007}}].

\bibitem{Petrosian:2004ft}
V.~Petrosian and S.-M. Liu, {\it {Stochastic Acceleration of Electrons and
  Protons. I. Acceleration by Parallel Propagating Waves}},  {\em Astrophys.
  J.} {\bf 610} (2004) 550--571,
  [\href{http://xxx.lanl.gov/abs/astro-ph/0401585}{{\tt astro-ph/0401585}}].

\bibitem{Simon:1985pz}
M.~{Simon}, W.~{Heinrich}, and K.~D. {Mathis}, {\it {Propagation of injected
  cosmic rays under distributed reacceleration}},  {\em Astrophys. J.} {\bf
  300} (1986) 32--40.

\bibitem{Seo:1994aa}
E.~S. {Seo} and V.~S. {Ptuskin}, {\it {Stochastic reacceleration of cosmic rays
  in the interstellar medium}},  {\em Astrophys. J.} {\bf 431} (1994) 705--714.

\bibitem{Liu:2004zi}
S.-M. Liu, V.~Petrosian, and F.~Melia, {\it {Electron Acceleration around the
  Supermassive Black Hole at the Galactic Center}},  {\em Astrophys. J.} {\bf
  611} (2004) L101--L104, [\href{http://xxx.lanl.gov/abs/astro-ph/0403487}{{\tt
  astro-ph/0403487}}].

\bibitem{Atoyan:2004ix}
A.~Atoyan and C.~D. Dermer, {\it {TeV emission from the Galactic Center
  black-hole plerion}},  {\em Astrophys.J.} {\bf 617} (2004) L123--L126,
  [\href{http://xxx.lanl.gov/abs/astro-ph/0410243}{{\tt astro-ph/0410243}}].

\bibitem{Scott:1975aa}
J.~S. {Scott} and R.~A. {Chevalier}, {\it {Cosmic-ray production in the
  Cassiopeia A supernova remnant}},  {\em Astrophys. J. Lett.} {\bf 197} (1975)
  L5--L8.

\bibitem{Cowsik:1984aa}
R.~Cowsik and S.~Sarkar, {\it {The evolution of supernova remnants as radio
  sources}},  {\em Mon. Not. R. Astron. Soc.} {\bf 207} (1984) 745.

\bibitem{Fan:2009kr}
Z.~Fan, S.~Liu, and C.~L. Fryer, {\it {Stochastic Electron Acceleration in the
  TeV Supernova Remnant RX J1713.7-3946: The High-Energy Cut-off}},  {\em Mon.
  Not. R. Astron. Soc.} {\bf 406} (2009) 1337--1349,
  [\href{http://xxx.lanl.gov/abs/0909.3349}{{\tt arXiv:0909.3349}}].

\bibitem{Mertsch:2011es}
P.~Mertsch and S.~Sarkar, {\it {Fermi gamma-ray `bubbles' from stochastic
  acceleration of electrons}},  {\em Phys.Rev.Lett.} {\bf 107} (2011) 091101,
  [\href{http://xxx.lanl.gov/abs/1104.3585}{{\tt arXiv:1104.3585}}].

\bibitem{Schlickeiser:2002pg}
R.~{Schlickeiser}, {\em Cosmic ray astrophysics}.
\newblock Springer, Berlin, 2002.

\bibitem{Melrose:1968aa}
D.~B. {Melrose}, {\it {The Emission and Absorption of Waves by Charged
  Particles in Magnetized Plasmas}},  {\em Ap\&SS} {\bf 2} (1968) 171--235.

\bibitem{Kulsrud:1969zz}
R.~Kulsrud and W.~P. Pearce, {\it {The Effect of Wave-Particle Interactions on
  the Propagation of Cosmic Rays}},  {\em Astrophys. J.} {\bf 156} (1969) 445.

\bibitem{Schlickeiser:1989aa}
R.~{Schlickeiser}, {\it {Cosmic-ray transport and acceleration. I - Derivation
  of the kinetic equation and application to cosmic rays in static cold
  media.}},  {\em Astrophys. J.} {\bf 336} (1989) 243--293.

\bibitem{Zhou:1990aa}
Y.~{Zhou} and W.~H. {Matthaeus}, {\it {Models of inertial range spectra of
  interplanetary magnetohydrodynamic turbulence}},  {\em J. Geophys. Res.} {\bf
  95} (1990) 14881--14892.

\bibitem{Dung:1990bb}
R.~{Dung} and R.~{Schlickeiser}, {\it {The influence of the Alfvenic cross and
  magnetic helicity on the cosmic ray transport equation. I - Isospectral slab
  turbulence}},  {\em Astron. Astrophys.} {\bf 240} (1990) 537--540.

\bibitem{Goldreich:1995zz}
P.~{Goldreich} and S.~{Sridhar}, {\it {Toward a theory of interstellar
  turbulence. 2: Strong alfvenic turbulence}},  {\em Astrophys. J.} {\bf 438}
  (1995) 763--775.

\bibitem{Cho:2002qi}
J.~Cho and A.~Lazarian, {\it {Compressible sub-alfvenic MHD turbulence in
  low-Beta plasmas}},  {\em Phys. Rev. Lett.} {\bf 88} (2002) 245001,
  [\href{http://xxx.lanl.gov/abs/astro-ph/0205282}{{\tt astro-ph/0205282}}].

\bibitem{Schlickeiser:1998zz}
R.~{Schlickeiser} and J.~A. {Miller}, {\it {Quasi-linear Theory of Cosmic-Ray
  Transport and Acceleration: The Role of Oblique Magnetohydrodynamic Waves and
  Transit-Time Damping}},  {\em Astrophys. J.} {\bf 492} (1998) 352.

\bibitem{Cho:2005mb}
J.~Cho and A.~Lazarian, {\it {Particle acceleration by MHD turbulence}},  {\em
  Astrophys. J.} {\bf 638} (2006) 811--826,
  [\href{http://xxx.lanl.gov/abs/astro-ph/0509385}{{\tt astro-ph/0509385}}].

\bibitem{Kaplan:1955aa}
S.~A. {Kaplan}, {\it {The Theory of the Acceleration of Charged Particles by
  Isotopic Gas Magnetic Turbulent Fields}},  {\em J. Exper. Theoret. Phys.}
  {\bf 2} (1956) 203--210.

\bibitem{Kardashev:1962aa}
N.~S. {Kardashev}, {\it {Nonstationarity of Spectra of Young Sources of
  Nonthermal Radio Emission}},  {\em Soviet~Ast.} {\bf 6} (1962) 317--327.

\bibitem{Park:1995aa}
B.~T. {Park} and V.~{Petrosian}, {\it {Fokker-Planck Equations of Stochastic
  Acceleration: Green's Functions and Boundary Conditions}},  {\em Astrophys.
  J.} {\bf 446} (1995) 699--716.

\bibitem{Becker:2006nz}
P.~Becker, T.~Le, and C.~D. Dermer, {\it {Time-Dependent Stochastic Particle
  Acceleration in Astrophysical Plasmas: Exact Solutions Including
  Momentum-Dependent Escape}},  {\em Astrophys.J.} {\bf 647} (2006) 539--551,
  [\href{http://xxx.lanl.gov/abs/astro-ph/0604504}{{\tt astro-ph/0604504}}].

\bibitem{Fedorov:2010zz}
Y.~{Fedorov} and M.~{Stehlik}, {\it {Stochastic acceleration by the induced
  electric field versus the Fermi acceleration}},  {\em Journal of Physics B
  Atomic Molecular Physics} {\bf 43} (2010) 185701.

\bibitem{Gull:1973aa}
S.~F. {Gull}, {\it {A numerical model of the structure and evolution of young
  supernova remnants}},  {\em Mont. Not. R. Astron. Soc.} {\bf 161} (1973)
  47--69.

\bibitem{Tammi:2008vg}
J.~Tammi and P.~Duffy, {\it {Particle-acceleration timescales in TeV blazar
  flares}},  {\em AIP Conf. Proc.} {\bf 1085} (2009) 475--478,
  [\href{http://xxx.lanl.gov/abs/0811.3573}{{\tt arXiv:0811.3573}}].

\bibitem{Schlickeiser:1989bb}
R.~{Schlickeiser}, {\it {Cosmic-Ray Transport and Acceleration. II. Cosmic Rays
  in Moving Cold Media with Application to Diffusive Shock Wave Acceleration}},
   {\em Astrophys. J.} {\bf 336} (1989) 264.

\bibitem{Murphy:1960aa}
G.~M. {Murphy}, {\em {Ordinary Differential Equations and Their Solutions}}.
\newblock D. Van Nostrand, New York, 1960.

\bibitem{Kamke:1977aa}
E.~{Kamke}, {\em {Differentialgleichungen: L\"osungsmethoden und L\"osungen, I,
  Gew\"ohnliche Differentialgleichungen}}.
\newblock B. G. Teubner, Leipzig, 1977.

\bibitem{Polyanin:1995aa}
A.~D. {Polyanin} and V.~F. {Zaitsev}, {\em {Handbook of Exact Solutions for
  Ordinary Differential Equations}}.
\newblock CRC Press, Boca Raton, 1995.

\end{thebibliography}
\end{document}